\documentclass[conference]{IEEEtran}
\IEEEoverridecommandlockouts
\usepackage{cite}
\usepackage{amsmath,amssymb,amsfonts}
\usepackage{algorithmic}
\usepackage{enumitem}
\usepackage{graphicx}
\usepackage{subfig}
\usepackage{textcomp}
\usepackage{xcolor}
\usepackage{mathtools}
\usepackage{booktabs} 
\usepackage{url}
\usepackage{fancyhdr}
\usepackage[]{mdframed}
\def\BibTeX{{\rm B\kern-.05em{\sc i\kern-.025em b}\kern-.08em
    T\kern-.1667em\lower.7ex\hbox{E}\kern-.125emX}}

\fancypagestyle{specialfooter}{%
  \fancyhf{}

  \fancyhead[L]{IEEE 7th International Conference on Big Data and Artificial Intelligence (BDAI 2024)}
  \fancyfoot[L]{\small{© 2024 IEEE. Personal use of this material is permitted.  Permission from IEEE must be obtained for all other uses, in any current or future media, including reprinting/republishing this material for advertising or promotional purposes, creating new collective works, for resale or redistribution to servers or lists, or reuse of any copyrighted component of this work in other works.}}
}

\begin{document}

\title{SecureNet: A Comparative Study of DeBERTa and Large Language Models for Phishing Detection\\
}

\author{\IEEEauthorblockN{Sakshi Mahendru}
\IEEEauthorblockA{
\textit{Palo Alto Networks}\\
California, USA \\
}
~\\
\and
\IEEEauthorblockN{Tejul Pandit}
\IEEEauthorblockA{
\textit{Palo Alto Networks}\\
California, USA \\
}


}

\maketitle

\begin{abstract}
Phishing, whether through email, SMS, or malicious websites, poses a major threat to organizations by using social engineering to trick users into revealing sensitive information. This not only compromises company's data security but also incurs significant financial losses. In this paper, we investigate whether the remarkable performance of Large Language Models (LLMs) can be leveraged for particular task like text classification, particularly detecting malicious content and compare its results with state-of-the-art Deberta V3 (DeBERTa using ELECTRA-Style Pre-Training with Gradient-Disentangled Embedding Sharing) model. We systematically assess the potential and limitations of both the approaches using comprehensive public datasets comprising diverse data sources such as email, HTML, URL, SMS, and  synthetic data generation. Additionally, we demonstrate how LLMs can generate convincing phishing emails, making it harder to spot scams and evaluate the performance of both models in this context. Our study delves further into the challenges encountered by DeBERTa V3 during its training phases, fine-tuning methodology and transfer learning processes. Similarly, we examine the challenges associated with LLMs and assess their respective performance. Among our experimental approaches, the transformer-based DeBERTa method emerged as the most effective, achieving a test dataset (HuggingFace phishing dataset) recall (sensitivity) of 95.17\% closely followed by GPT-4 providing a recall of 91.04\%. We performed additional experiments with other datasets on the trained DeBERTa V3 model and LLMs like GPT 4 and Gemini 1.5. Based on our findings, we provide valuable insights into the effectiveness and robustness of these advanced language models, offering a detailed comparative analysis that can inform future research efforts in strengthening cybersecurity measures for detecting and mitigating phishing threats.
\end{abstract}


\section{Introduction} \label{Intro}
\thispagestyle{specialfooter}
The rapid advancement of internet technology has significantly exacerbated the risks associated with email phishing. Consequently, the threat of falling victim to such attacks has also increased \cite{paper_ref1}. It is imperative for organizations to build reliable and robust email filtering systems to minimize errors and address problems related to oversight and false positives.


Innovative solutions leveraging cutting-edge technologies are needed to bolster cybersecurity defenses against phishing attacks. Large language models (LLMs) such as GPT-4\cite{openai2024gpt4} have revolutionized natural language processing, enabling the generation of highly convincing human-like text. However, these advancements have also empowered malicious actors to craft sophisticated phishing emails at scale. Recent studies have shown that LLMs can generate phishing websites and emails that convincingly imitate well-known brands while deploying evasive tactics to elude anti-phishing systems\cite{phishbots}.

This study addresses these challenges by comparing phishing detection mechanisms leveraging state-of-the-art LLMs (GPT-4 and Gemini 1.5\cite{geminiteam2024gemini}) alongside DeBERTa V3\cite{he2023debertav3}. Our approach tackles the diverse and evolving nature of phishing attacks across various communication channels, such as email, SMS, URLs, and webpages. We compile a comprehensive dataset from diverse sources to capture a wide spectrum of phishing domains. Additionally, we augment our dataset with synthetic data generated using GPT-4, allowing us to thoroughly explore the capabilities of LLMs and DeBERTa V3 in identifying phishing attempts across different categories.

Through empirical and comparative analysis, we evaluate the performance of these models in mitigating phishing risks, safeguarding organizational reputations, and reducing costs associated with cyberattacks. We also delve into the vulnerabilities of each technique, providing valuable insights essential for making informed security decisions and enhancing cybersecurity strategies.

The study is organized as follows: Section \ref{dataDesc}, Dataset Description, details the datasets used, their sources, and their characteristics across various communication channels. Section \ref{methodology}, Methodology, provides an overview of the methodologies employed, including the use of LLMs and DeBERTa V3. Section \ref{expSetup}, Experimental Setup, outlines the model training, validation, and testing processes. Section \ref{results}, Results, presents the findings and comparative analysis of model performance. Section \ref{relatedWork} presents a thorough Literature Survey, reviewing existing research and frameworks on phishing detection mechanisms. Section \ref{futureScope} explores the future scope of the research, and Section \ref{conclusion} concludes by summarizing the key findings and their implications for enhancing cybersecurity strategies.

\section{Dataset Description} \label{dataDesc}
As part of our research into ways to detect phishing attempts, we utilized a diverse set of datasets that were specifically chosen to encompass a broad range of potential phishing scenarios and legitimate communications. To ensure the accuracy of the data, each dataset underwent a rigorous pre-processing phase, which involved removing any null, empty, duplicate and invalid entries. Furthermore, we analyzed the data to identify patterns that could help us gain a deeper understanding of the information. The details of each of the data sources and their corresponding characteristics are provided in the following Sections.

\subsection{Dataset Details}
\subsubsection{\textbf{HuggingFace Phishing Dataset}\cite{hfdataset}}\mbox{}\\
The HuggingFace phishing dataset represents a rich compilation of phishing datasets sourced from reputable repositories and tailored specifically for classification and phishing detection tasks. This dataset comprises a heterogeneous mix of textual samples, including URLs, SMS messages, email messages, and HTML code. Each record within this dataset is meticulously labeled as either 1 (indicating phishing) or 0 (indicating benign), providing a foundational framework for our experimentation.

The datasets encompass a compilation sourced from four distinct origins.
\begin{itemize}
    \item The Mail Dataset comprises 18,000 Enron employee emails, aiding in understanding phishing email patterns.
    \item The SMS Message Dataset, with over 5,971 texts, facilitates classifying malicious and legitimate messages.
    \item An extensive URL Dataset of 800,000 URLs, including 52\% legitimate and 47\% phishing domains, assists in studying URL-based phishing attacks.
    \item The Website Dataset contains 80,000 websites, including 50,000 legitimate and 30,000 phishing sites, providing insights into web-based phishing threats.
\end{itemize}

These datasets offer a comprehensive approach to phishing detection across various communication channels, enhancing the study's robustness and applicability.

The dataset underwent preprocessing to ensure balance among classes and phishing sources. However, a disproportionate abundance of URLs (97\%) posed a risk of bias, potentially misrepresenting real-world scenarios. To address this, a balanced subset, ``combined\_reduced," was created, incorporating varied data types (emails, SMS, websites). Neglecting these could lead to the model overlooking crucial features and nuances. The ``combined\_reduced" corpus aimed to mitigate bias, ensuring a comprehensive representation of phishing sources across different communication channels. This balanced approach enhances the dataset's suitability for training models to detect phishing attempts effectively.

\subsubsection{\textbf{Nazario and Nigerian Fraud Dataset}\cite{champa2024curated}}\mbox{}\\
The Nazario\_5 and Nigeria\_5 datasets, sourced from a reputable open-access repository, play a pivotal role in our investigation. These datasets have been curated to provide a comprehensive collection of both legitimate and phishing-related communications, thereby enriching our dataset with diverse and representative samples.

\begin{enumerate}[label=(\roman*)]
    \item \underline{Nazario\_5 dataset}

    It is constructed by amalgamating a diverse array of legitimate emails from renowned datasets such as SpamAssasin.csv\cite{spamAss}, TREC Spam Corpus (TREC\_05.csv, TREC\_06.csv, TREC\_07.csv)\cite{trecCorpus}, and CEAS\_08.csv\cite{ceasData}, with the Nazario.csv\cite{nazario} dataset. The details of each dataset is provided below.
    \begin{itemize}
        \item SpamAssassin: A total corpus of 6047 emails with a 31\% contribution of spam messages.
        \item TREC Spam Corpus: The dataset is curated by applying an automated spam filter on a sequence of emails which are then evaluated by humans to create a benchmark dataset.
        \item CEAS\_08: Corpus from a Spam Challenge held in 2008 used for laboratory evaluations.
        \item Nazario: Publicly available collection of 4558 phishing emails.
    \end{itemize}

    This process ensures that the dataset encompasses a wide range of email communication styles and content types, including plain text, HTML, and attachments. Additionally, the inclusion of emails from multiple sources enhances the diversity and richness of the dataset, enabling a more comprehensive evaluation of phishing detection mechanisms.

    \item \underline{Nigeria\_5 dataset}

    It combines emails from the SpamAssasin.csv, TREC Spam Corpus (TREC\_05.csv, TREC\_06.csv, TREC\_07.csv), and CEAS\_08.csv with the Nigerian\_Fraud.csv\cite{nigeria} dataset, thereby further enriching our dataset with authentic and phishing-related communications. Fraudulent emails often contain deceptive information intended to persuade recipients to send large sums of money to the sender. One of the most well-known types of these scams is the Nigerian Letter, also known as ``419" Fraud. The Nigerian fraud dataset is built with the idea of capturing and analyzing such deceptive emails to improve detection and prevention techniques which contributes to 2500 fraud emails ranging from 1998 to 2007. We aim to capture unique linguistic and contextual cues indicative of phishing attempts, thereby assessing their impact on the effectiveness of our phishing detection mechanisms.
\end{enumerate}

We provide detailed analysis on the dataset composition and the trends observed in Section \ref{nnChar}. It helps to highlight the evolving patterns of phishing emails over the years, providing valuable insights into the dataset's dynamics.

\subsubsection{\textbf{Synthetic Data Generation}}\mbox{}\\
The motivation for utilizing LLMs for synthetic data generation stems from dual objectives.

\begin{enumerate}[label=(\Alph*)]
    \item  We conducted experiments to assess the potential misuse of LLMs for crafting phishing emails with jailbreaking techniques, confirming concerns raised by previous studies \cite{li2023multistep, shen2024do}. Our findings highlight the ease with which attackers can exploit LLM capabilities for large-scale phishing attacks, leveraging deceptive tactics to extract sensitive information \cite{greshake2023youve}..
    
    \item Utilizing the advanced language model GPT-4, we generated synthetic data to enrich our dataset with diverse phishing attempts, including social engineering and deception tactics. This augmentation strategy enhances the variability and realism of our dataset, enabling a more robust evaluation of phishing detection mechanisms.
\end{enumerate}

Email samples generated using GPT-4 accompanied by an explanation provided by the LLM, offering insights into why the email falls into a specific category are all added in the Github repository\footnote{\url{https://github.com/tejulp/PhishingDetection}}. We have also added our prompts used to construct phishing emails for each of the category (i.e. legitimate and phishing) in the same repository for reference.

By amalgamating these diverse datasets, we aimed to construct a comprehensive and representative corpus that encapsulates the multifaceted nature of phishing attacks and benign communications. This holistic approach ensures that our experimentation framework is sufficiently robust and capable of accurately evaluating the efficacy of our phishing detection mechanisms across a myriad of scenarios and data distributions. Moreover, the detailed description and characterization of each dataset offered valuable insights into the challenges and complexities of phishing detection in real-world settings, thereby advancing our understanding of cybersecurity threats and defenses. Detailed characteristics of each dataset used can be found in Section \ref{dataChar}.

\subsection{Dataset Characteristics} \label{dataChar}

\subsubsection{\textbf{HuggingFace Phishing Dataset}}\mbox{}\\
The dataset consists of two columns - text and label. Each entry in the `text' column represents one of the following forms of communication:  URL, HTML text, SMS message or Email content. The `label' column assigns a classification label to each entry, indicating whether it is a phishing attempt or benign communication. After initial pre-processing, there are a total of 77,568 samples in the dataset, evenly distributed among the different categories. To facilitate DeBERTa V3’s fine-tuning process, the dataset is split into three sets: training, testing, and validation. We adopt a 70-30 split, allocating 70\% of the data for training and the remaining 30\% for testing and validation. The latter subset is then divided equally into validation and test sets. 

It was also imperative to ensure that the classes 0 (benign text) and 1 (phishing samples) are balanced in the overall corpus as well as in each of the individual sets. Table \ref{tab:hfDataset} presents the distribution of labels across the dataset and its corresponding subsets.

 \begin{table}[htbp]
    \caption{Class Distribution - HuggingFace phishing dataset}
    \begin{center}
    \begin{tabular}{|c|c|c|c|c|c|c|c|}
    \hline
    \multicolumn{2}{|c|}{\textbf{Complete Dataset}} & \multicolumn{2}{|c|}{\textbf{Training Set}} & \multicolumn{2}{|c|}{\textbf{Validation Set}} & \multicolumn{2}{|c|}{\textbf{Test Set}} \\
    \hline
    Class & Class & Class & Class & Class & Class & Class & Class \\
    0 & 1 & 0 & 1 & 0 & 1 & 0 & 1 \\
    \hline
    44975 & 32702 & 31433 & 22864 & 6761 & 4874 & 6672 & 4964\\
    \hline
    \end{tabular}
    \label{tab:hfDataset}
    \end{center}
\end{table}

Furthermore, to enhance the dataset's organization and utility, each sample was augmented with an additional column termed ``category." This column delineated the type of content present in the text column. Specifically, emails and SMS messages were categorized under the ``text" category, while URLs were designated as ``url," and HTML content was categorized as ``web." This categorization facilitated efficient data management and streamlined the analysis process.

\subsubsection{\textbf{Nazario and Nigerian Fraud Dataset}}\label{nnChar}\mbox{}\\
The individual datasets (Nazario\_5 and Nigeria\_5) were concatenated, resulting in a dataframe with a total shape of (9396, 7). The columns included in this dataframe are sender email, receiver email, subject, body, date, label, and URLs (found within the email body). Following the initial preprocessing step, which involved removing null, empty, and duplicate entries, the dataset was reduced to 7382 samples. Subsequently, any rows containing invalid entries in the subject or body, identified by the presence of the string \emph{``=?utf-8?”}, were removed, resulting in a final sample size of 5330. From this dataset, a subset of 800 rows was selected for evaluating the selected models. The distribution of data across the relevant labels is provided in Table \ref{tab:nnDataset}. Notably, the dataset is equally distributed among labels to ensure a fair evaluation process

\begin{table}[htbp]
    \caption{Class Distribution - Nazario and Nigerian dataset}
    \begin{center}
    \begin{tabular}{|c|c|c|c|}
    \hline
    \multicolumn{2}{|c|}{Complete Dataset} & \multicolumn{2}{|c|}{Sampled Dataset} \\
    \hline
    Class 0 & Class 1 & Class 0 & Class 1 \\
    \hline
    2598 & 2732 & 408 & 392 \\
    \hline
    \end{tabular}
    \label{tab:nnDataset}
    \end{center}
\end{table}

Given that the corpus comprises a combination of phishing email datasets sourced from diverse origins, it is essential to consider the temporal aspect of the data. The datasets vary in the timeframe during which the phishing emails were observed, reflecting different periods of email-based cyber threats.

\begin{figure}
        \centerline{\includegraphics[width=8cm]{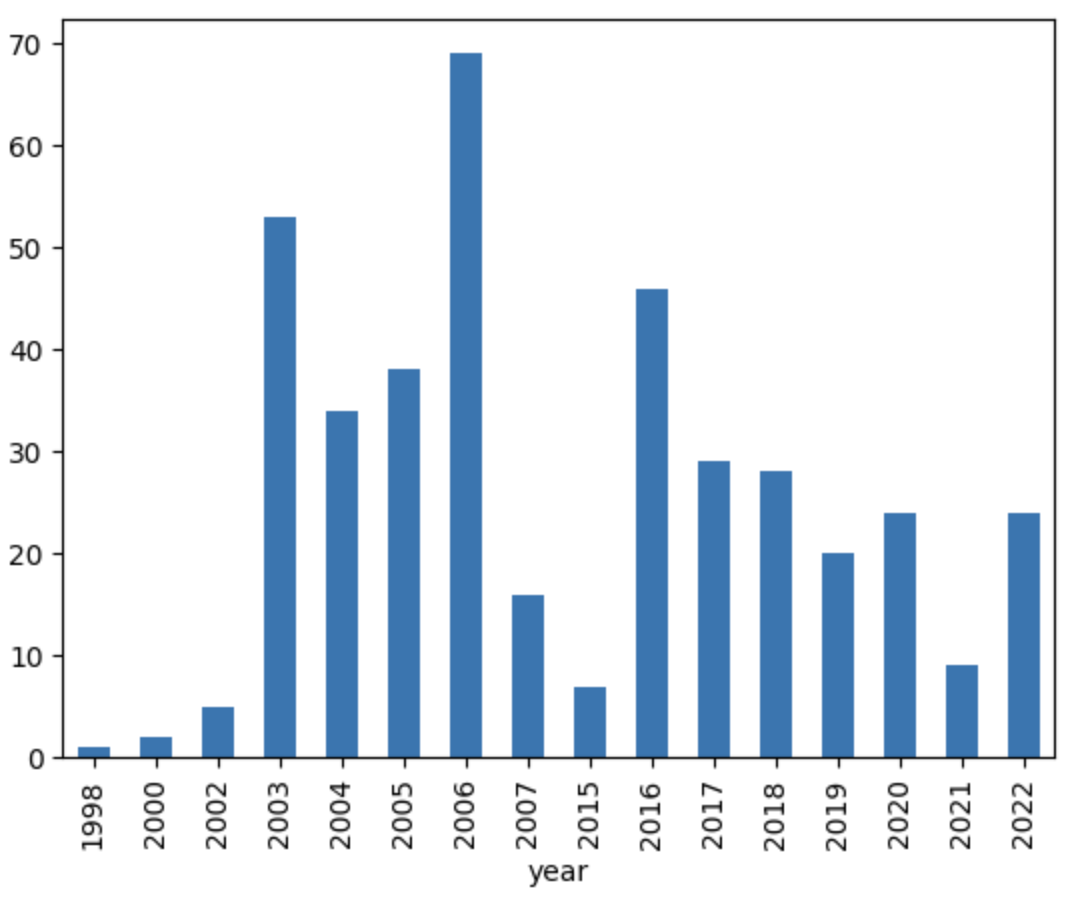}}
        \caption{Temporal Distribution on sample set}
        \label{fig:tempAnalysis}
\end{figure}

\mbox{}\\
\underline{Temporal Analysis}\mbox{}\\
The dataset includes a date column for each row, facilitating temporal analysis of phishing threats. Visual analysis from 1998 to 2022 as seen in Fig. \ref{fig:tempAnalysis} reveals distinct trends, notably pre and post-2007. These trends may stem from variations in dataset collection methods. A deeper linguistic analysis was conducted to enhance understanding of phishing tactics' evolution over time.



\begin{figure}%
    \centering
    \subfloat[]{{\includegraphics[width=8cm]{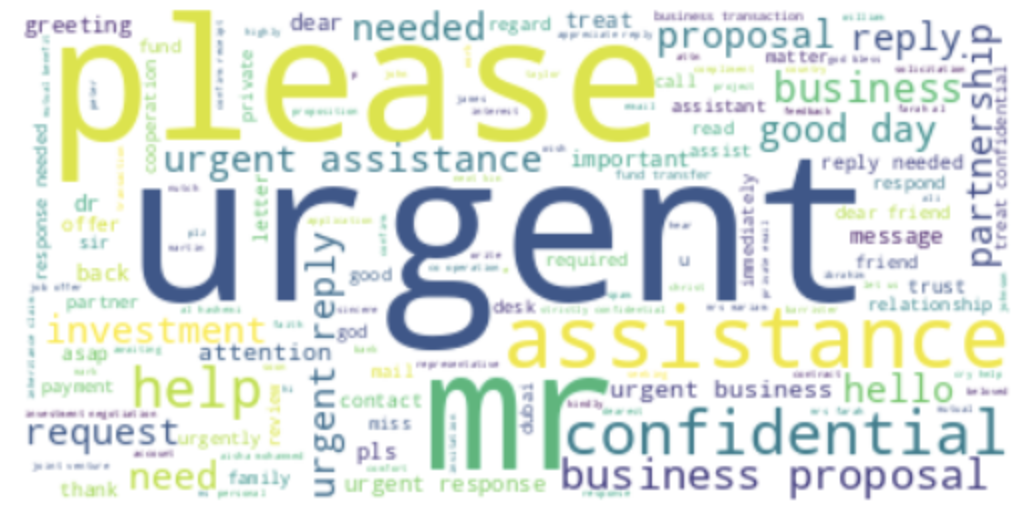} \label{fig:pre2007} }}%
    \qquad
    \subfloat[]{{\includegraphics[width=8cm]{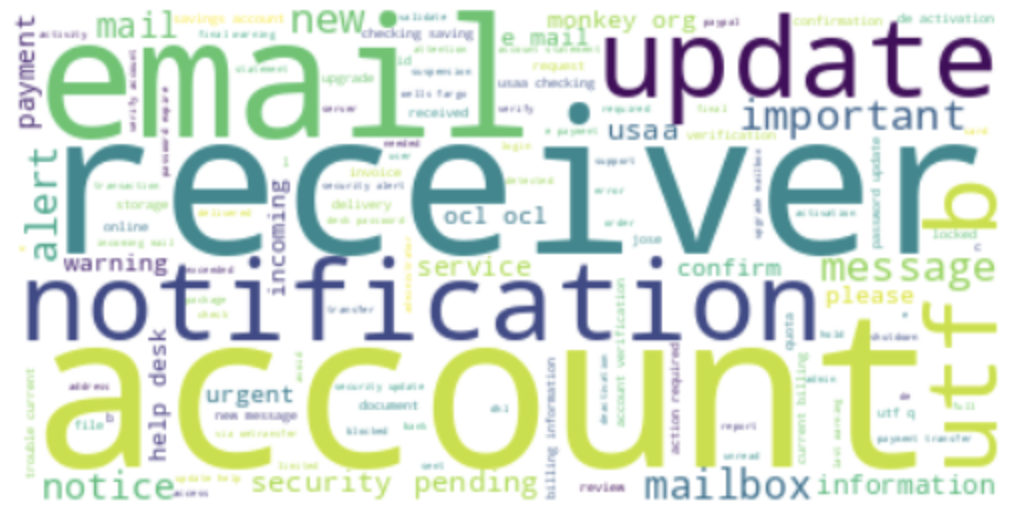} } \label{fig:post2007}}%
    \caption{(a) Word cloud for pre-2007 email corpus, (b) Word cloud for post-2007 email corpus}%
    \label{fig:wcloud}%
\end{figure}

\mbox{}\\
\underline{Liguistic Trends Analysis}\mbox{}\\
To delve deeper into our hypothesis concerning the evolution of phishing tactics over time, we conducted a thorough analysis by dividing the sample corpus into two distinct subsets: pre-2007 and post-2007. Word clouds in Figures \ref{fig:pre2007} and \ref{fig:post2007} visually depict prevalent terminologies used in phishing emails during each timeframe. Pre-2007 (Fig. \ref{fig:pre2007}) shows terms related to urgent business matters, suggesting a focus on fake urgency tactics. Post-2007 (Fig. \ref{fig:post2007}) exhibits a shift towards account notifications and technological terms, reflecting the rise of technology. This evolution highlights cybercriminals' adaptability to exploit technological advancements. Through the analysis of linguistic characteristics in phishing emails across various timeframes, valuable insights emerge regarding the evolving strategies employed by cybercriminals to deceive users and circumvent cybersecurity measures. 



\subsubsection{\textbf{Synthetic Data Generation}}\label{synDataChar}\mbox{}\\
Initially, generating both phishing and legitimate email content posed challenges with LLMs when utilizing prompt injection.
These included - 
\begin{enumerate}
    \item For phishing email, we observed that LLMs tend to produce placeholder content like [Your Company's IT Security Team], generate concise emails, and predominantly focus on the security domain. 
    \item There was a consistent presence of urgency in the generated phishing emails, serving as a hint.
\end{enumerate}

To address these issues, we iteratively refined our approach and eventually developed instructions that were effective at curbing inherent behaviour and guiding the model to generate phishing and legitimate emails meeting our criteria. Subsequently, all the generated samples underwent manual grading and verification to ensure accuracy and reliability.

The resulting dataset consisted of 40 samples signifying legitimate email samples and 80 samples of phishing emails. 

\begin{table}[htbp]
    \caption{Class Distribution - Synthetic dataset}
    \begin{center}
    \begin{tabular}{|c|c|c|c|c|c|}
    \hline
    \multicolumn{2}{|c|}{Complete} & \multicolumn{2}{|c|}{Training-Validation-Test} &  \multicolumn{2}{|c|}{Test}\\
    \multicolumn{2}{|c|}{Dataset} & \multicolumn{2}{|c|}{Dataset} &  \multicolumn{2}{|c|}{Dataset}\\
    \hline
    Class 0 & Class 1 & Class 0 & Class 1 & Class 0 & Class 1 \\
    \hline
    39 & 80 & 20 & 40 & 19 & 40 \\
    \hline
    \end{tabular}
    \label{tab:synDataset}
    \end{center}
\end{table}

Further, we partitioned this dataset nearly evenly for training DeBERTa V3. One portion was designated for the training, validation, and testing process of the DeBERTa model, while the remaining subset was reserved for independent testing with synthetic data. The distribution of data across these sets is summarized in Table \ref{tab:synDataset}. To ensure each subset contained a representative selection of synthetic samples, all samples within the training-validation-test dataset were categorized as ``synthetic." Further details regarding experimentation with synthetic data are provided in Section \ref{sec1deb}.

\mbox{}\\
\underline{Prompt Design}\mbox{}\\
The prompt for phishing emails was structured to adhere to specific guidelines aimed at mimicking real-world phishing tactics such as impersonating a brand or high-profile individual, instilling a sense of fake urgency, or simply generating false notifications regarding updates/account activities requiring recipients to log in.

In contrast, the prompt for legitimate emails was tailored to produce samples that mimic suspicious email characteristics while remaining legitimate and harmless to recipients. The instructions within the prompt emphasized the importance of generating samples that serve as effective training materials for educating employees on identifying potential phishing attempts and distinguishing them from benign communications.

\section{Methodology} \label{methodology}

In this section, we assess the comparative effectiveness of DeBERTa V3, against other large language models, namely OpenAI's GPT-4 and Google's Gemini Pro, both utilizing chain-of-thought (COT) and instruction prompting methodologies. By investigating the capabilities and nuances of these models, our objective is to gauge their efficacy in distinguishing phishing emails from legitimate messages. We employ specific evaluation metrics, such as precision, recall, accuracy, and F1-score, to quantify the performance of these models. Furthermore, our study provides insights for both approaches to phishing content detection within contemporary cyber-security landscapes, thereby addressing a pertinent challenge in the field.

\subsection{\textbf{DeBERTa V3 Model as a Text Classifier}}
We employed the DeBERTa V3  model, a state-of-the-art transformer-based language model renowned for its superior performance in natural language processing tasks. DeBERTa represents a significant advancement over foundational models such as BERT, RoBERTa, and ELECTRA, integrating their core architectures while implementing additional enhancements to improve model effectiveness \cite{tranformerTop}. Unlike previous models, DeBERTa addresses the limitations of token positioning awareness, ensuring a more nuanced understanding of token context and improving overall classification accuracy \cite{debertaRev}. Furthermore, the introduction of DeBERTa V3 builds upon the original model by replacing masked language modeling (MLM) with replaced token detection (RTD), a more sample-efficient pre-training task \cite{debertaTok}. This enhancement contributes to setting a new state-of-the-art benchmark among models with similar structures, further enhancing the classification performance of DeBERTa \cite{debertaTok}.

We conducted our experiment on 'DeBERTa-v3-base,' for both tokenization and training the classifier. The DeBERTa V3 base model consists of 12 layers with a hidden size of 768. With only 86 million backbone parameters and a vocabulary of 128,000 tokens, the model introduces 98 million parameters in the Embedding layer. Training of this model utilized a dataset of 160GB, consistent with the approach taken for DeBERTa V2. We provide a comprehensive elaboration of the experimental setup for DeBERTa V3 in Section \ref{expSetup}.


\subsection{\textbf{LLM with prompt tuning as a Text Classifier}} \label{llmClassifier}

For our experiment, we chose two models, GPT-4 and Gemini 1.5, due to their notable advancements in the latest generative AI. These models showcase extensive capabilities in processing large datasets effectively \cite{geminiBlog} and are renowned for their achievements across various downstream tasks. 

\mbox{}\\
\underline{Prompt Design}

Prompting involves providing carefully crafted instructions to a pre-trained model, enabling it to bridge the gap between its inherent capabilities and the desired performance level. To ensure consistency in evaluating diverse data sources like HTML/WEB, URL, SMS, and Email, we devised a unified prompt inspired by \cite{koide2024chatspamdetector}.

Our approach involved prompting the model to justify its decisions and assign a score ranging from 1 to 10, reflecting its reasoning capability. A score of ``1" suggests a low probability of the email being a phishing attempt, indicating legitimacy, while a score of ``10" indicates a high likelihood of phishing, showcasing multiple characteristics associated with phishing. This process of reasoning and scoring proved invaluable for refining and enhancing our prompts iteratively, while also facilitating result analysis.

We utilized the Chain of Thought (COT) prompting technique guided by a set of rules. The integration of this rule-based methodology ensured that the model followed a specific evaluation strategy and conducted comprehensive analysis across different communication channels. This method encourages the model to engage in step-by-step thinking and analysis, eliciting critical evaluation and reasoning abilities, resulting in improved performance across various datasets and significantly higher efficacy compared to standard prompts. 

The prompt systematically evaluated inputs against following rules.
\begin{itemize}
    \item Rule 1: Analyze emails for language, sender authenticity, and formatting.
    \item Rule 2: Scrutinize URLs for anomalies and verify destination links.
    \item Rule 3: Beware of suspicious SMS messages and verify sender identity and links.
    \item Rule 4: Inspect HTML code for hidden threats and irregularities.
    \item Rule 5: Verify website security and legitimacy before sharing personal data.
    \item Rule 6: Be cautious with email attachments, especially from unknown sources.
    \item Rule 7: Double-check inputs, apply critical thinking, and rectify any issues for improved phishing detection.
\end{itemize}

Lastly, the prompt specified returning a JSON response with three key fields: `is\_phishing', indicating the predicted label by the LLM model; `reason', which outlines the thought process and rationale; and `score', used to represent multiple characteristics associated with phishing.




\section{Experimental Setup} \label{expSetup}

\subsection{\textbf{DeBERTa V3 Model as a Text Classifier}}

In this section, we delved into the fine-tuning process of the DeBERTa V3 model. We outlined all the challenges DeBERTa V3 faced during both training and transfer learning processes. Section \ref{sec1deb} explores the intricacies of preparing the training dataset for fine-tuning purposes. Section \ref{sec2deb} shifts the focus towards the fine-tuning of hyperparameters. Section \ref{sec3deb} assessed the model’s effectiveness across various datasets mentioned above in Section \ref{dataDesc}.

\subsubsection{Training Dataset Preparation for DeBERTa V3 fine-tuning}\label{sec1deb}

The process of preparing the training dataset for fine-tuning DeBERTa V3 involved meticulous steps to ensure the model's proficiency in classifying various types of phishing texts from diverse sources. Initially, we recognized the necessity for a representative dataset capable of capturing a wide spectrum of phishing instances, irrespective of their origin. Our first experiment with the Nazario and Nigerian Fraud dataset revealed limitations in model performance due to its narrow focus on email-based attacks. Transitioning to the HuggingFace phishing dataset provided a more comprehensive training corpus, covering various communication channels such as email, SMS, URL, and HTML. Despite improvements, the model struggled to generalize when evaluated on synthetic data. To address this, we integrated a subset of synthetic data into the training set, resulting in our best-performing model. This iterative process highlighted the importance of dataset diversity in enhancing the model's ability to detect phishing texts across different sources. By combining real-world and synthetic data, we achieved a more robust and versatile phishing detection model, emphasizing the significance of dataset representation in training machine learning models for cybersecurity applications.

\subsubsection{Hyperparameter Fine-Tuning for DeBERTa Model} \label{sec2deb}

Fine-tuning the DeBERTa V3 model's hyperparameters was crucial for optimizing its performance for our problem statement and dataset. We adjusted the learning rate, selecting $1e-7$ to balance model stability and convergence speed. Experimentation with the warmup ratio led us to choose 0.01, facilitating gradual learning rate adjustment for improved convergence. Weight decay was tuned to 0.01 to control overfitting while capturing meaningful patterns. Most experiments used a fixed number of epochs (i.e., 5), but the final experiment extended training to 10 epochs to assess convergence behavior. These adjustments aimed to optimize the model's performance for phishing detection on our dataset.

\subsubsection{Training Trends  Observations} \label{sec3deb}

This section outlines training trends observed across three different training configurations: \\
DeBERTa\_1: Utilizing the HuggingFace dataset alone.\\
DeBERTa\_2: Combining the HuggingFace dataset with synthetic data for 5 epochs.\\
DeBERTa\_3: Extending the training duration to 10 epochs for the same dataset as DeBERTa\_2. 

\begin{itemize}
    \item \underline{DeBERTa\_1}\\
    Training exclusively on the HuggingFace phishing dataset served as a foundational reference point for assessing model performance. This approach enabled the model to familiarize itself with a diverse array of real-world phishing examples. Fig. \ref{fig:chart1} illustrates the observed training and validation trends over 5 epochs, corresponding to the hyperparameters outlined in Section \ref{sec2deb}.\\
    \begin{figure}
            \centerline{\includegraphics[width=8cm]{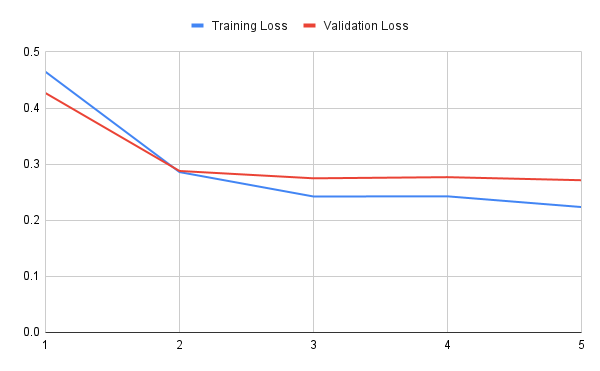}}
            \caption{Training and Validation curve for HuggingFace Dataset}
            \label{fig:chart1}
    \end{figure}
    \item \underline{DeBERTa\_2}\\
    DeBERTa\_1's ability to generalize to synthetic phishing instances, was observed to be suboptimal. Integrating synthetic data alongside the HuggingFace dataset for 5 epochs resulted in enhanced model performance. This amalgamation broadened the training data spectrum, exposing the model to a diverse array of phishing patterns and bolstering its capacity to extrapolate to unseen data. We illustrate the overall training trend in Fig. \ref{fig:chart2}, demonstrating improvement compared to the preceding method.
    \begin{figure}
            \centerline{\includegraphics[width=8cm]{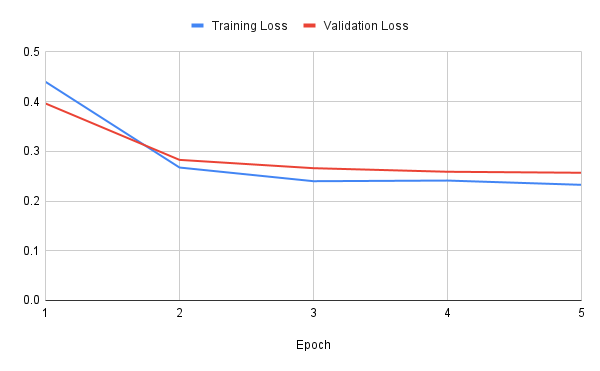}}
            \caption{Training and Validation curve for 5 epochs on HuggingFace and Synthetic Dataset}
            \label{fig:chart2}
    \end{figure}
    \item \underline{DeBERTa\_3}\\
    Extending the training duration to 10 epochs for the combined HuggingFace and synthetic dataset resulted in further improvements in model performance. The additional training iterations allowed the model to refine its representations and better capture the nuances of phishing texts. Our claims are supported by observing the training and validation loss chart provided in Fig. \ref{fig:chart3}.
    \begin{figure}
        \centerline{\includegraphics[width=8cm]{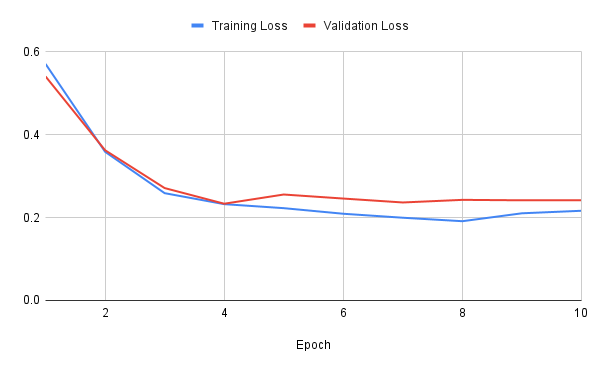}}
        \caption{Training and Validation curve for 10 epochs on HuggingFace and Synthetic Dataset}
        \label{fig:chart3}
\end{figure}
\end{itemize}

\subsection{\textbf{LLM with prompt tuning as a Text Classifier}}

In this section, we utilize LLMs for text classification. 

We employed LLM and a prompt as explained in Section \ref{llmClassifier}. The GPT-4 (specifically, GPT-4 Turbo preview) model was utilized for evaluating the Nazario and Nigerian and HuggingFace phishing datasets, while for the synthetic dataset, Gemini 1.5 was employed to avoid biases inherited from the source model (GPT-4) while generating the synthetic data. This cross-evaluation allowed for a more comprehensive understanding of the strengths and weaknesses of the models being assessed. The temperature setting for both the models is set to 0.1.

Each sample from the datasets used for evaluation consists of two fields, namely, text and label. The text value encompasses various types of input such as HTML, URLs, SMS, or emails, while the label is binary, denoted as either 0 or 1. 0 signifies a ``safe email," indicating it's a ham, while 1 denotes a ``phishing content," indicating a spam. The prediction response from LLM consists of a JSON response as explained in Section \ref{llmClassifier} wherein the ``is\_phishing" value is the predicted label to compare against the actual label.




\section{Results} \label{results}

The experimental results from our diverse setup configurations provide valuable insights into the performance of SOTA models, LLMs and DeBERTa, across diverse experimental setups and datasets. We present a comprehensive analysis of the results obtained from each experimental setup, focusing on the metrics used for evaluation.

\subsection{Evaluation Metrics}

The metrics used include recall (sensitivity to capture most phishing attempts), precision (accuracy in predicting phishing attempts), F1 score (harmonic mean of precision and recall for balanced performance) and accuracy (overall correct classifications). These metrics collectively evaluate the model's effectiveness in phishing detection.

In a spam classifier scenario, the primary objective would be to optimize recall to minimize the chances of missing any spam emails. Conversely, maximizing precision becomes crucial when the priority is to avoid mistakenly classifying significant non-spam emails as spam. Therefore, \emph{the choice between maximizing recall or precision depends on the specific objectives and priorities of the classifier's application}.

\subsection{Comparative Study}

We conduct a comparative analysis of Large Language Models against three variants of Deberta V3. Each model is evaluated based on their performance in distinguishing between legitimate and phishing emails.


The focus of this study is to build a highly robust and effective system to detect phishing email. Therefore, we focused primarily on the recall scores for our comparative analysis.

\begin{figure}
    \centerline{\includegraphics[width=8cm]{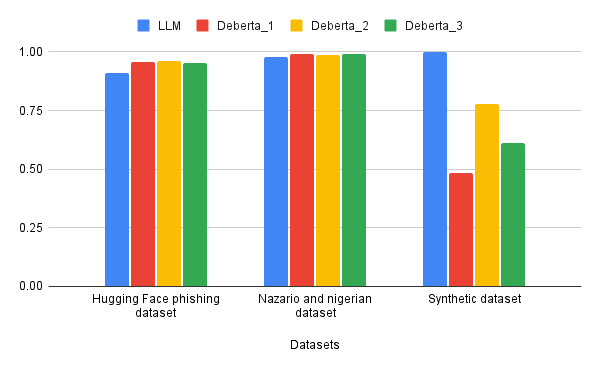}}
    \caption{Recall scores comparison}
    \label{fig:cmp1}
\end{figure}

Recall measures the proportion of actual phishing emails that are correctly identified as such by the model. Given the critical nature of identifying as many phishing emails as possible to prevent potential security breaches, maximizing recall is paramount. We present the recall scores for LLM and the three variants of the DeBERTa across all the test datasets. in Fig. \ref{fig:cmp1}. This comparison allows us to assess each model's ability to correctly identify phishing emails while minimizing false negatives. 

The findings depicted in Fig. \ref{fig:cmp1} emphasize that while LLM and DeBERTa exhibit consistently similar performance levels on the public dataset, LLM notably outperforms DeBERTa on the synthetic dataset. It's essential to acknowledge that DeBERTa models outperform LLM on public datasets, likely due to the representative nature of the training dataset captured. Moreover, the DeBERTa model demonstrated significantly improved performance on synthetic data following its training with a subset thereof, underscoring the pivotal role of representative datasets in optimizing model efficacy.

\begin{figure}
    \centerline{\includegraphics[width=8cm]{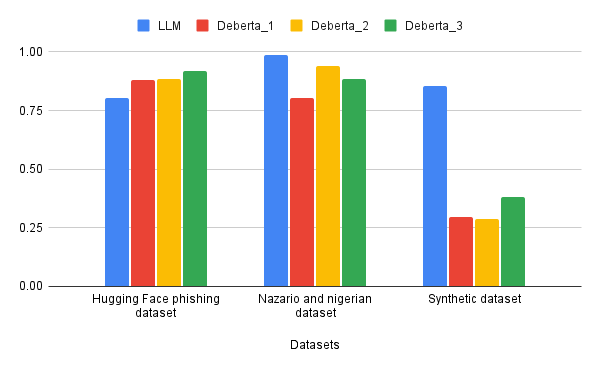}}
    \caption{F1-scores comparison}
    \label{fig:cmp2}
\end{figure}

Additionally, we present a comparison of the F1-score for all four models in Fig. \ref{fig:cmp2}. The F1-score provides a balanced measure of a model's precision and recall, offering insights into its overall performance in phishing detection. The findings illustrated in Fig. \ref{fig:cmp2}, align with the observations drawn from the recall score comparison, with one notable exception in the synthetic dataset. Despite the DeBERTa\_3 model exhibiting a lower recall than the DeBERTa\_2 model, it achieves a higher F1-score. This discrepancy suggests an overall improved performance in terms of both recall and precision for the DeBERTa\_3 model in this context. As a result of a better or competitive F1-score on almost all datasets, we consider DeBERTa\_3 as our best performing DeBERTa V3 model.

\begin{figure}%
    \centering
    \subfloat[]{{\includegraphics[width=8cm]{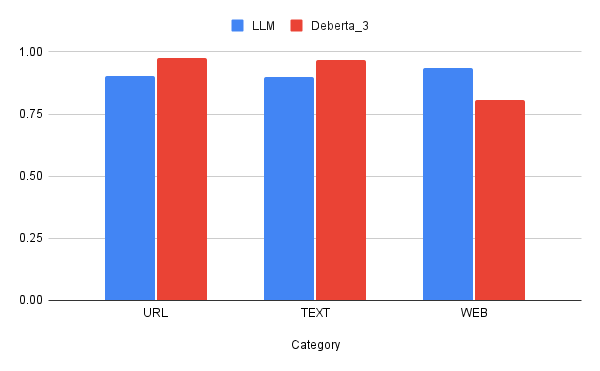} \label{fig:cmp3} }}%
    \qquad
    \subfloat[]{{\includegraphics[width=8cm]{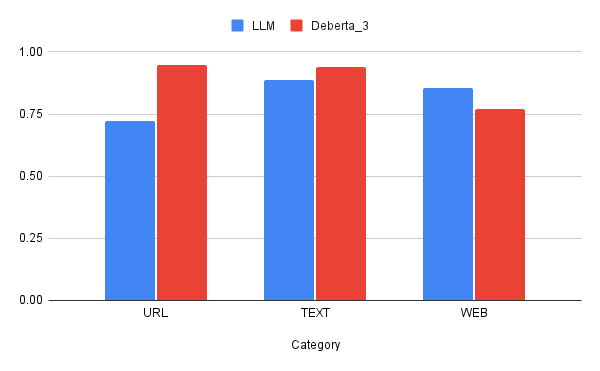} } \label{fig:cmp4}}%
    \caption{(a) Recall scores comparison between LLM and DeBERTa\_3, (b) F1-scores comparison between LLM and DeBERTa\_3}%
    \label{fig:wcloud}%
\end{figure}

Expanding the inclusion of synthetic data samples for refining the DeBERTa model, coupled with an increase in the number of epochs, yields enhanced performance, particularly evident in the synthetic corpus. Moreover, we conducted an in-depth investigation to compare the efficacy of LLM with DeBERTa\_3 across various categories within the HuggingFace dataset, including text (comprising SMS and emails), URLs, and web/HTML pages. The comparison of recall and F1 scores is visually represented in Fig. \ref{fig:cmp3} and Fig. \ref{fig:cmp4}, offering insights into the models' performance across distinct data categories.

Upon examining Fig. \ref{fig:cmp3} and Fig. \ref{fig:cmp4}, a noteworthy observation emerges: while LLM and DeBERTa demonstrate competitive scores in the text category, DeBERTa exhibits superior performance in identifying URLs, whereas LLMs excel in detecting web/HTML pages.

The observed performance disparity could be attributed to the inherent strengths and weaknesses of each model architecture. DeBERTa, known for its robustness in handling sequential data, may possess an advantage in accurately identifying URLs, which typically exhibit a distinct pattern and structure. Conversely, LLMs, renowned for their contextual understanding and reasoning capabilities, might excel in deciphering the nuanced context and content within web/HTML pages, which often contain intricate textual and visual elements.

This performance discrepancy underscores the importance of selecting the appropriate model architecture based on the specific characteristics and complexities of the data domain. Furthermore, it highlights the potential synergies that could be leveraged by integrating multiple models or refining existing architectures to achieve more comprehensive and accurate detection capabilities across diverse data categories.

Moreover, Table \ref{tab:cmpDataAll} provides a complete view of all the metric scores computed and puts forth a comparison between LLM and DeBERTa\_3 model across all the datasets used for evaluation. In our opinion, including more samples from synthetic data for the DeBERTa V3 model training can potentially help in improving the Recall with a corresponding overall improvement in F1-score. 

\begin{table*}[t]
    \centering
    \caption{LLM v/s DeBERTa\_3 model metrics comparison}
    \begin{tabular}{|c|c|c|c|c|c|c|c|c|}
    \hline
     & \multicolumn{2}{|c|}{\textbf{Recall}} & \multicolumn{2}{|c|}{\textbf{Precision}} & \multicolumn{2}{|c|}{\textbf{Accuracy}} & \multicolumn{2}{|c|}{\textbf{F1-score}} \\
    \hline
    \textbf{Dataset} & LLM & DeBERTa\_3 & LLM & DeBERTa\_3 & LLM & DeBERTa\_3 & LLM & DeBERTa\_3 \\ 
    \hline
    HuggingFace &	0.91045 &	0.95176 &	0.72189 &	0.88579 &	0.82067 &	0.88650 &	0.80528 &	0.91759 \\
    Phishing Dataset &  &  &  &  &  &  &  & \\
    \hline
    Nazario \& &	0.97980	& 0.99051	& 0.98980 &	0.79847 &	0.79129 &	0.93980 &	0.98477 &	0.88418 \\
    Nigerian dataset &  &  &  &  &  &  &  & \\
    \hline
    Synthetic Dataset	& 1.00000 &	0.61111 &	0.74766 &	0.27500 &	0.77311 &	0.28570	& 0.85561	& 0.37931\\
    \hline
    \end{tabular}
    \label{tab:cmpDataAll}
\end{table*}

Lastly, an important observation regarding model efficiency was the average latency per sample taken by LLMs compared to DeBERTa models for prediction as depicted in Table \ref{tab:latData}. Despite DeBERTa models requiring longer time for fine-tuning, their prediction time is significantly faster—approximately 3,000 times faster than that of LLMs for the same sample. This substantial difference in latency highlights the practical advantages of using DeBERTa models in real-time phishing detection systems, where quick response times are critical for mitigating threats effectively. The speed advantage of DeBERTa models makes them more suitable for deployment in high-throughput environments where timely detection and response to phishing attempts are paramount.

\begin{table}[htbp]
    \caption{Latency Comparison - LLM v/s DeBERTa}
    \begin{center}
    \begin{tabular}{|c|c|}
    \hline
    \textbf{Models} & \textbf{Average latency/sample}\\
    \hline
    LLM & 13-15 seconds \\
    \hline
    DeBERTa & 3.75 milliseconds \\
    \hline
    \end{tabular}
    \label{tab:latData}
    \end{center}
\end{table}

\subsection{LLM Outcome Analysis}
We conducted an in-depth analysis of the LLM using COT prompting, offering insights into its reasoning process. Notably, the LLM shows a very high score of 8 or 9 in false positives and a very low score of 0 in false negatives. This implies that LLMs tend to exhibit binary outcomes, showcasing a notable propensity for categorizing content either as phishing or legitimate, with limited intermediate classifications. In the subsequent subsections we cover some of the top categories identified from false positives and false negatives where LLM encounters challenges.

\subsubsection{False Positives}
\begin{enumerate}[label=(\roman*)]
    \item \textbf{URL Analysis}: LLMs often misclassify legitimate URLs as suspicious due to:
    \begin{enumerate}[label=(\alph*)]
        \item LLMs bring in their additional knowledge of brands in the world due to which certain urls that were not directly linked to known brands were incorrectly tagged as potential phishing attempts. For example, a URL redirecting through `www.newsisfree.com' but claiming affiliation with CNN is flagged due to lack of brand association. Similarly, ``https://www.rokuki.com/family-of-discovery-go-channels-launch-on-roku/" was misclassified due to failure in recognizing ``Rokuki" as a legitimate brand and the LLM assumed it to be an erroneous association to the popular brand Roku.
        \item LLMs struggle to discern the legitimacy of urls with multiple subdomains, such as ``https://annrulesweblog.wordpress.com/", which is genuine WordPress blogs is considered to be mimicking and hence wrongly tagged as phishing.
    \end{enumerate}
    \item \textbf{Email/SMS Analysis}: Reasons for misclassification in emails or sms by LLM include:
    \begin{enumerate}[label=(\alph*)]
        \item It was observed that despite some of the emails being genuine, due to a tone of urgency incorporated in the requests were flagged as phishing by LLM.
        \item LLMs would simply categorize absence of personalized greeting or signature in email or sms as phishing.
        \item An interesting observation was when a false assumption was made by LLMs for emails with significantly old dates such as ``2002-09-26" which were considered to be suspicious under the pretext of being copied or templated phishing attempts.
    \end{enumerate}
\end{enumerate}

\subsubsection{False Negatives}
\begin{enumerate}[label=(\roman*)]
    \item It was observed in some cases that certain factors influenced the decision made by LLMs towards legitimate classification causing a deviation from other indicators of potential phishing. One such example was of a survey email that allowed the option of opting out which was a positive sign but the LLM neglected other relevant details such as sender's email address or domain authenticity hindering conclusive analysis.
    \item Links within legitimate services may still pose risks; for instance, requests like "To be removed from our distribution lists, please click here..." may hide malicious intent which were not captured by LLMs.
    \item Urls that lack a secure http protocol but have the brand name associated accurate and devoid of any misspellings, LLMs fail to find a phishing attempt.
\end{enumerate}

\section{Literature Survey} \label{relatedWork}

Research for email phishing using machine learning and deep learning methods began as early as 2015. Many supervised learning algorithms\cite{janezmartino2024classifying} have set the baseline for detection. \cite{mlp_phishing} proposed PDMLP, a new method proving better results than most traditional machine learning techniques.  Combining these methods improved robustness and prediction performance \cite{textmining_phishing}, emphasizing the importance of Natural Language Processing (NLP) methodologies for phishing analysis. The paper focuses on extracting relevant features from email headers, sender email addresses, content, and attachments, such as, URL domains, HTML code, and attachment file types. However, traditional machine learning models struggle to capture the complex patterns and intricate dependencies present in sequential data, such as email content and headers. 

Deep learning architectures, particularly those for sequential data, have revolutionized phishing detection. Recurrent Neural Networks (RNNs) \cite{schmidt2019recurrent}, Long Short-Term Memory (LSTM) networks \cite{lstm}, and Transformers \cite{vaswani2023attention} have shown promising results in detecting subtle anomalies. \cite{nguyen2018deep} introduces hierarchical LSTM networks (H-LSTMs) for detecting sophisticated phishing attempts. Various other deep learning-based methods \cite{nguyen2024pioneering, chatterjee2019deep, Nmachi_2023}, help to predict malicious emails with high confidence through evidence. 

Attention-based architectures \cite{han2021transformer} enhance interpretability by capturing linguistic patterns and contextual relationships in email data. Transfer learning from pre-trained models like DeBERTa \cite{devlin2019bert} fine-tunes large-scale datasets for phishing detection, making it more efficient and effective. DeBERTa incorporates enhancements such as learned positional embeddings and dynamic masking, surpassing traditional BERT models. Large language models like GPT-4 \cite{han2021pretrained} offer promising results without specific training data for downstream tasks. Despite their strengths in natural language understanding and generation, they have limitations \cite{liu2023lost}.

In phishing detection, Large Language Models (LLMs) are extensively utilized. \cite{koide2024chatspamdetector} introduces ChatSpamDetector, employing GPT-4 to detect scams, outperforming baseline systems. \cite{nahmias2024prompted} utilizes an ensemble of LLMs to detect phishing emails, achieving superior results. \cite{uddin2024explainable} presents a transformer-based model, DistilBERT, with 98.48

LLMs are proficient in identifying phishing or scam email signs, as shown by \cite{jiang2024detecting}. However, \cite{phishbots} demonstrates LLMs' ability to create convincing phishing content. \cite{heiding2023devising} illustrates competitive LLMs' dual role as both spam attackers and detectors, increasing incentives for spear-phishing attacks. These advancements underscore the need for new countermeasures.

Existing studies on phishing detection have been limited in their scope, failing to conduct comprehensive comparative analyses across various communication channels such as email, SMS, URLs, and webpages using Bert models and large language models like GPT-4. Our work addresses this gap by presenting detailed analysis.

\section{Future Work} \label{futureScope}

Future research in phishing detection could focus on optimizing DeBERTa V3 configurations and fine-tuning LLMs like GPT-4 for specific tasks, enhancing their precision and recall. Exploring advanced synthetic data utilization by creating datasets that better mimic real-world scenarios can further improve model training and evaluation. Additionally, integrating multimodal data (text, images, and metadata) could provide a more comprehensive understanding of email content, leading to more accurate phishing detection. These advancements aim to enhance the efficacy and robustness of language models in real-world applications, ensuring more reliable and precise detection of phishing emails.

\section{Conclusion} \label{conclusion}
In this study, we evaluated the capabilities of state-of-the-art language models in detecting phishing emails, focusing on DeBERTa and large language models (LLMs) like GPT-4 and Gemini 1.5. Our analysis highlighted their strengths and limitations, emphasizing the importance of aligning training and evaluation data to enhance model performance. DeBERTaV3 demonstrated improved generalization and accuracy with properly aligned data, while LLMs excelled in detecting phishing within synthetic datasets, indicating their potential to recognize phishing patterns even in newly emerging tactics, providing a dynamic and adaptive approach to cybersecurity. Additionally, LLMs' ability to reason and extract suspicious entities enhances phishing detection mechanisms, offering a comprehensive and dynamic approach to mitigating cyber threats.

\bibliographystyle{plain}
\bibliography{phishingNew.bib} 

\end{document}